\documentclass[epj,final,fleqn]{svjour}

\usepackage{amssymb,amstext,graphicx}

\begin{document}
\title{Enskog-Landau kinetic equation for multicomponent mixture.\\ 
Analytical calculation of transport coefficients}

\titlerunning{Analytical calculation of transport coefficients}


\author{A.E. Kobryn\and M.V. Tokarchuk\and Y.A. Humenyuk}

\authorrunning{A.E. Kobryn {\slshape et al.}}

\institute{Institute for Condensed Matter Physics of the National Academy of 
Sciences of Ukraine,\\ 1~Svientsitskii St., UA-79011 Lviv, Ukraine;
\email{alex@icmp.lviv.ua}}



\date{\today}

\abstract{%
The Enskog-Landau kinetic equation is considered to describe non-equilibrium 
processes of a mixture of charged hard spheres. This equation has been 
obtained in our previous papers by means of the non-equilibrium statistical 
operator method. The normal solution of this kinetic equation found in the 
first approximation using the standard Chapman-Enskog method is given. On 
the basis of the found solution the flows and transport coefficients have 
been calculated. All transport coefficients for multicomponent mixture of 
spherical Coulomb particles are presented analytically for the first time. 
Numerical calculations of thermal conductivity and thermal diffusion 
coefficient are performed for some specific mixtures of noble gases of high 
density. We compare the calculations with those ones for point-like neutral 
and charged particles.
}


\maketitle

Construction of kinetic equations for dense gases and plasma is one of the 
most important problems in the kinetic theory of classical systems. A 
sequential kinetic theory of dense systems does not exist yet. The 
Enskog-Landau kinetic equation has been obtained recently in \cite{1} to 
describe transport processes in non-equilibrium system of charged hard 
spheres. This equation has its name due to the structure of total collision 
integral. This integral contains terms of revised Enskog theory, kinetic 
mean field theory and Landau-like collision integral \cite{2}. The influence 
of the last term on system behaviours has been of our main interest. This 
term is caused by the long-range interactions in a system. In particular, it 
was shown \cite{2} that in the case of small densities and weak interactions 
the force autocorrelation function and the entire last term in the total 
collision integral converts to the usual Landau collision integral for a 
rarefied plasma.

The new kinetic equation and its collision integral are adequate only for 
systems which can be modelled by char\-ged hard spheres. The great credit 
for this result is shared by a choice of interparticle interaction potential 
in ad\-di\-ti\-ve-li\-ke form: hard spheres interaction plus certain 
``smo\-oth'' long-range part (Coulomb interaction in our case). It allowed 
to avoid divergency at short distances. Unfortunately, a logarithmic 
divergency at long distances still remains, and to eliminate it one should 
introduce a cut-off radius of integration (like a Debye one). But unlike the 
classical Debye formula, in this case we used a modified one, which takes 
into account particle sizes $\sigma$. For the Enskog-Landau kinetic equation 
\cite{1} a normal solution has been found by means of the standard 
Chapman-Enskog method. A stress tensor $\tens{\Pi}$ and heat flow vector 
$\bf q$ have been obtained as well. Expressions for transport coefficients 
like bulk $\kappa$ and shear $\eta$ viscosities and thermal conductivity 
$\lambda$ have been derived from structures of $\tens{\Pi}$ and $\bf q$. 
Numerical calculation of transport coefficients for neutral and once-ionized 
argon shows a good agreement between the theory and experimental data. In 
\cite{3,4} these results were generalized to {\sl non-stationary} 
non-equilibrium process. Whereas to find the normal solution the 
Chapman-Enskog method \cite{5} is used in \cite{1,2}, the much more powerful 
method of boundary conditions \cite{6} is used in \cite{3,4}. In the 
limiting case of a stationary non-equilibrium process, the results of 
\cite{4} completely convert to those of \cite{1}. For hydrodynamic 
description of fast processes it is better to use the method of boundary 
conditions \cite{6}.

Application of the theory to a multicomponent system was performed 
step-by-step. The Enskog-Landau kinetic equation for $M$-component 
($M$$\,\geqslant\,$$2$) mixture of charged hard spheres has been proposed in 
\cite{7}. Just the same, the normal solution, flows and transport 
coefficients have been found by means of the standard Chapman-Enskog method 
for a two-component system only. New transport coefficients which appear in 
multicomponent systems are mutual diffusion $D^{\alpha\beta}$ and thermal 
diffusion $D^\alpha_{\rm T}$ coefficients (here $\alpha$,$\beta$ are mixture 
indices). Numerical calculation of the obtained transport coefficients 
showed a good agreement between the developed theory, experimental data, 
results of other theories and MD simulations.

In view of identifying of a normal solution of the kinetic equation for a 
multicomponent system of charged particles results of \cite{7} are not the 
first. Namely, it is worthy to note that the normal solution of the kinetic 
equation for completely ionized plasma is found in \cite{8} using the 
standard Chapman-Enskog method in the 3rd approximation. But dense systems 
of finite size particles are consistently considered for the first time in 
our papers \cite{1,2,3,4,7}.

In this letter we present our solution of the Enskog-Landau kinetic equation 
for a multicomponent mixture of charged hard spheres. Similarly to \cite{7}, 
we use the standard Chapman-Enskog method. It is known \cite{5} that the 
correction expression for one-particle distribution function in the first 
approximation can be chosen in two different ways. According to the first 
one described in \cite{9}, the correction for one-particle distribution 
function of $\alpha$-kind is proportional to ${\mathbf d}_\alpha$ -- a 
diffusion thermodynamic force of $\alpha$-kind only. The second way is 
proposed in \cite{5}. In this case the correction is proportional to a 
certain superposition of diffusion thermodynamic forces of all components of 
a mixture. It was shown \cite{10} that the second method gives much better 
results because, unlike the first method, after crossing to linear 
thermodynamics equations it is important that Onsager's reciprocal relations 
obey. Nevertheless, in \cite{7} for a two-component system we used the first 
method because the complication (or generalization) like in \cite{5} is 
essential for two-component systems in some cases only. Namely, when the 
density of particle number of some mixture component (or components) is not 
conserved. Such a situation can be realized in gas mixtures where chemical 
reactions between components may take place, or in multicomponent mixtures, 
where transitions between states with different internal degree of freedoms 
are possible.

Let us consider the Enskog-Landau kinetic equation for multicomponent 
mixture of charged hard spheres:
\begin{equation}
\left[\frac{\partial}{\partial t}+{\mathrm i}L(1_\alpha)\right]
	f_1(x_1^\alpha;t)=\sum_{\beta=1}^M
	I_{\mathrm{coll}}\Big(f_1(x_1^\alpha),f_1(x_2^\beta)\Big),
	\label{e1}
\end{equation}
here $f_1$ denotes the one-particle distribution function, 
$x\equiv\{{\mathbf r};{\mathbf p}\}$ is a set of phase coordinates in a 
phase space while ${\mathbf r}$ and ${\mathbf p}$ denote the Cartesian 
coordinate and particle momentum, respectively. Collision integral of this 
equation has an additive structure
\begin{equation}
I_{\mathrm{coll}}=I_{\mathrm{HS}}^{(0)}+I_{\mathrm{HS}}^{(1)}+
I_{\mathrm{MF}}+I_{\mathrm{L}}\label{e2}
\end{equation}
and each term in (\ref{e2}) is defined like in \cite{7}: the first and  
second ones are from so-called hard sphere part of interparticle 
interaction -- collision integral of the revised Enskog theory RET:
\begin{eqnarray*}
&&I_{\mathrm{HS}}^{(0)}=\int\mathrm{d}\mathbf{v}_2^\beta\mathrm{d}\varepsilon
\mathrm{d}b\;b|\text{\slshape\bfseries g}^{\alpha\beta}|
\text{\slshape g}_2^{\alpha\beta}(\sigma_{\alpha\beta}|n,\beta)
\times{}\\
&&\left(f_1(\mathbf{r}_1^\alpha,\mathbf{v}_1^\alpha{}')
f_1(\mathbf{r}_2^\beta,\mathbf{v}_2^\beta{}')-
f_1(\mathbf{r}_1^\alpha,\mathbf{v}_1^\alpha)
f_1(\mathbf{r}_2^\beta,\mathbf{v}_2^\beta)\right),\\\\
&&I_{\mathrm{HS}}^{(1)}=\sigma_{\alpha\beta}^3
\int\mathrm{d}\hat{\mathbf{r}}_{12}^{\alpha\beta}\mathrm{d}\mathbf{v}_2^\beta
\theta(\hat{\mathbf{r}}_{12}^{\alpha\beta}\cdot
\text{\slshape\bfseries g}^{\alpha\beta})\times{}\\
&&\left[\hat{\mathbf{r}}_{12}^{\alpha\beta}
\text{\slshape g}_2^{\alpha\beta}
(\mathbf{r}_{12}^{\alpha\beta}|n,\beta)\times{}\right.\\
&&\left(f_1(\mathbf{r}_1^\alpha,\mathbf{v}_1^\alpha{}')
\mbox{\boldmath$\nabla$}
f_1(\mathbf{r}_2^\beta,\mathbf{v}_2^\beta{}')-
f_1(\mathbf{r}_1^\alpha,\mathbf{v}_1^\alpha)\mbox{\boldmath$\nabla$}
f_1(\mathbf{r}_2^\beta,\mathbf{v}_2^\beta)\right)+{}\\
&&\frac12\left(\hat{\mathbf{r}}_{12}^{\alpha\beta}\cdot
\mbox{\boldmath$\nabla$}
\text{\slshape g}_2^{\alpha\beta}(\mathbf{r}_{12}^{\alpha\beta}|n,\beta)
\right)\times{}\\
&&\left.\left(f_1(\mathbf{r}_1^\alpha,\mathbf{v}_1^\alpha{}')
f_1(\mathbf{r}_2^\beta,\mathbf{v}_2^\beta{}')+
f_1(\mathbf{r}_1^\alpha,\mathbf{v}_1^\alpha)
f_1(\mathbf{r}_2^\beta,\mathbf{v}_2^\beta)\right)\right];
\end{eqnarray*}
the third one is caused by taking into account of long-range interparticle 
interaction in the mean field approximation KMFT (this term is of the first 
order in interaction):
\begin{eqnarray*}
I_{\mathrm{MF}}&=&\frac{1}{m_\alpha}\int\mathrm{d}\mathbf{r}_2^\beta
\frac{\partial}{\partial\mathbf{r}_1^\alpha}\mathit{\Phi}^{\mathrm{l}}
(|\mathbf{r}_{12}^{\alpha\beta}|)\times{}\\
&&\text{\slshape g}_2^{\alpha\beta}(\mathbf{r}_1^\alpha,\mathbf{r}_2^\beta;t)
n_1(\mathbf{r}_2^\beta;t)\frac{\partial}{\partial\mathbf{v}_1^\alpha}
f_1(x_1^\alpha;t);
\end{eqnarray*}
and, finally, the third one is the so-called Landau-like collision term (it 
is of the second order in interaction):
\begin{eqnarray*}
&&I_{\mathrm{L}}=\frac{1}{4(m^*)^2}
\frac{\partial}{\partial\mathbf{v}_1^\alpha}\int\mathrm{d}x_2\;
\text{\slshape g}_2(\mathbf{r}_1^\alpha,\mathbf{r}_2^\beta;t)
\left[
\frac{\partial\mathit{\Phi}^{\mathrm{l}}(|\mathbf{r}_{12}^{\alpha\beta}|)}
{\partial\mathbf{r}_{12}^{\alpha\beta}}\right]\times{}\\
&&\int\limits_{-\infty}^0\!\mathrm{d}t'\!
\left[\frac{\partial\mathit{\Phi}^{\mathrm{l}}
(|\mathbf{r}_{12}^{\alpha\beta}+\text{\slshape\bfseries g}t'|)}
{\partial\mathbf{r}_{12}^{\alpha\beta}}\right]\!\!
\left[\frac{\partial}{\partial\mathbf{v}_1^\alpha}-
\frac{\partial}{\partial\mathbf{v}_2^\beta}\right]
f_1(x_1^\alpha;t)f_1(x_2^\beta;t).
\end{eqnarray*}
In expressions for $I_{\mathrm{HS}}^{(0)}$, $I_{\mathrm{HS}}^{(1)}$, 
$I_{\mathrm{MF}}$ and $I_{\mathrm{L}}$ we used the following designations: 
$b$ -- impact parameter, $\beta$ -- an analogue of local inversed 
temperature, $\varepsilon$ -- azimuthal angle of scattering, 
$\text{\slshape\bf g}^{\alpha\beta}$ -- relative velocity of $\alpha$- and 
$\beta$-kind particles, $\text{\slshape g}_2^{\alpha\beta}$ -- two-particle 
correlation function, $m^*$ -- reduced mass, $m_\alpha$ -- partial masses of 
particles, $n$ -- total density of particles number, $n_\alpha$ -- partial 
densities of particles numbers, $\hat{\mathbf{r}}_{12}^{\alpha\beta}$ -- 
unit vector along $\mathbf{r}_{12}^{\alpha\beta}$ direction, $\mathbf{v}'$ 
-- velocities of hard spheres after collision, $\theta(\ldots)$ -- Heaviside 
unit step function.

In the limit of a system of point-like charged particles of low density, the 
collision integral (\ref{e2}) transforms to the usual Landau collision 
integral \cite{11,12}. Following a concept of consistent description of 
kinetics and hydrodynamics of non-equilibrium processes 
\cite{13,14,15,16,17}, kinetic equation (\ref{e1}) should be solved together 
with local conservation laws \cite{4} for additive invariants. These 
additive invariants in collisions (or scattering) of charged hard spheres 
are mass (or total density), momentum and total energy \cite{5,9,11,12}. It 
should be noted that for rarefied systems it was sufficient to consider 
kinetic energy only, while in dense systems potential interaction energy is 
essential and cannot be neglected. 

The solution of Equation (\ref{e1}) found in the first approximation by 
means of the Chapman-Enskog method is
\begin{equation}
f_1^{(1)}(x_1^\alpha;t)=f_1^{(0)}(x_1^\alpha;t)\Big[1+
	\varphi(x_1^\alpha;t)\Big],\label{e3}
\end{equation}
where $f_1^{(0)}(x_1^\alpha;t)$ is the local quasi-equilibrium Maxwell 
one-particle distribution function:
\begin{eqnarray*}
&&f_1^{(0)}(x_1^\alpha;t)=\\
&&n_\alpha({\mathbf r}_1;t)\left[\frac{m_\alpha}
	{2\pi k_{\mathrm B}T({\mathbf r}_1;t)}\right]^{3/2}
\exp\left\{-\frac{m_\alpha\left(c_1^\alpha({\mathbf r}_1;t)\right)^2}
	{2k_{\mathrm B}T({\mathbf r}_1;t)}\right\}.
\end{eqnarray*}
This function is the solution of Equation (\ref{e1}) in the zeroth 
approximation and satisfies the Fredgolm conditions. Correction 
$\varphi(x_1^\alpha;t)$ reads:
\begin{eqnarray}
\varphi(x_1^\alpha;t)&=&-
	A_\alpha[(C_1^\alpha)^2]{\mathbf C}_1^\alpha({\mathbf r}_1;t)
	\cdot\mbox{\boldmath$\nabla$}\ln T({\mathbf r}_1;t)
	\label{e4}\\
&&-B_\alpha[(C_1^\alpha)^2]\left[{\mathbf C}_1^\alpha{\mathbf C}_1^\alpha
	-\frac13(C_1^\alpha)^2\tens{I}\right]:\mbox{\boldmath$\nabla$}
	{\mathbf V}({\mathbf r}_1;t)\nonumber\\
&&+n\sum_{\beta=1}^ME_{\alpha\beta}[(C_1^\alpha)^2]
	{\mathbf C}_1^\alpha({\mathbf r}_1^\alpha;t)\cdot
	{\mathbf d}_\beta({\mathbf r}_1;t).\nonumber
\end{eqnarray}
Here
\[
{\mathbf C}^\alpha={\mathbf C}^\alpha({\mathbf r};t)=\left[\frac{m_\alpha}
	{2k_{\mathrm B}T}\right]^{1/2}\!{\mathbf c}^\alpha({\mathbf r};t),\;
	{\mathbf c}^\alpha({\mathbf r};t)={\mathbf v}^\alpha-
	\langle{\mathbf v}\rangle.
\]
In other words, $\langle{\mathbf v}\rangle$ is nothing but hydrodynamical 
ve\-lo\-ci\-ty ${\mathbf V}({\mathbf r};t)$. Functionals 
$A_\alpha[(C_1^\alpha)^2]$, $B_\alpha[(C_1^\alpha)^2]$, 
$E_{\alpha\beta}[(C_1^\alpha)^2]$ are defined by the Sonine-Laguerre 
polynomials \cite{5,9}.

Having the solution of Equation (\ref{e1}) in the first approximation, one 
can calculate the stress tensor $\tens{\Pi}$, heat flow vector ${\mathbf 
q}$ and diffusion velocity ${\mathbf V}^{\mathrm d}$ in the same 
approximation. The expression for $\tens{\Pi}$ reads:
\[
\tens{\Pi}=p\tens{I}-\kappa\left(\mbox{\boldmath$\nabla$}
	\cdot{\mathbf V}\right)\tens{I}-2\eta\tens{S}, 
\]
where $p$ is total pressure in the first approximation ($\tens{I}$ is the 
unit tensor, $\tens{S}$ is the velocities shift tensor):
\begin{eqnarray*}
p=nk_{\mathrm B}T&+&\frac{2}{3}\pi k_{\mathrm B}T\sum_{\alpha,\beta=1}^M
	\sigma_{\alpha\beta}^3
	\text{\slshape g}_2^{\alpha\beta}n_\alpha n_\beta-{}\\
&&\frac{2}{3}\pi\sum_{\alpha,\beta=1}^Mn_\alpha n_\beta
	\int\limits_{\sigma_{\alpha\beta}}^{\infty}{\mathrm d}x\;
	x^3\text{\slshape g}_2^{\alpha\beta}(x)\frac{\partial}{\partial x}
	\Phi_{\alpha\beta}^{\mathrm l}(x),
\end{eqnarray*}
($\text{\slshape g}_2^{\alpha\beta}(x)$ is the binary quasi-equilibrium 
correlation function, $\Phi_{\alpha\beta}^{\mathrm l}(x)$ is a long-range 
potential of interaction); $\kappa$ is the bulk viscosity of a 
multicomponent mixture:
\begin{equation}
\kappa=\frac{4}{9}\sum_{\alpha,\beta=1}^M
	\sigma_{\alpha\beta}^4\text{\slshape g}_2^{\alpha\beta}
	n_\alpha n_\beta\sqrt{2\pi k_{\mathrm B}T\mu_{\alpha\beta}}=
	\sum_{\alpha,\beta=1}^M\kappa_{\alpha\beta},\label{e5}
\end{equation}
$\eta$ is the shear viscosity of a multicomponent mixture:
\begin{eqnarray}
\eta&=&\frac{3}{5}\kappa+\frac{1}{2}k_{\mathrm B}T\sum_{\alpha=1}^M 
	n_\alpha B_\alpha(0)+{}\label{e6}\\
&&\frac{2}{15}\pi k_{\mathrm B}T
	\sum_{\alpha,\beta=1}^M\sigma_{\alpha\beta}^3
	\text{\slshape g}_2^{\alpha\beta}\mu_{\alpha\beta}n_\alpha n_\beta
	\left[\frac{B_\alpha(0)}{m_\alpha}+\frac{B_\beta(0)}{m_\beta}\right].
	\nonumber
\end{eqnarray}
The expression for heat flow ${\mathbf q}$ reads:
\[
{\mathbf q}=-\lambda\mbox{\boldmath$\nabla$}T+
	\sum_{\alpha=1}^M\omega_\alpha{\mathbf d}_\alpha.
\]
Quantities $\omega_\alpha$ are connected with a matter transfer due to a 
temperature gradient (the Soret effect) and due to a heat transfer caused by 
a gradient of concentration (the Dufour effect). If one also takes into 
account a barrodiffusion process, this constitutes the total contribution 
into a heat flow from cross transfer processes. $\lambda$ is the thermal 
conductivity coefficient of a multicomponent mixture:
\begin{eqnarray}
\lambda&=&\sum_{\alpha,\beta=1}^M
	\frac{3k_{\mathrm B}\;\kappa_{\alpha\beta}}{m_\alpha+m_\beta}-
	\sqrt{2k_{\mathrm B}^3T}\times{}\label{e7}\\
&&\left[\frac{5}{4}\sum_{\alpha=1}^M
	\frac{n_\alpha}{\sqrt m_\alpha}
	\left[A_\alpha(1)-A_\alpha(0)\right]+{}\right.\nonumber\\
&&\left.\frac{2\pi}{3}\sum_{\alpha,\beta=1}^M
	\frac{\sigma_{\alpha\beta}^3\text{\slshape g}_2^{\alpha\beta}
	n_\alpha n_\beta}{m_\alpha+m_\beta}
	\left[\frac{3\mu_{\alpha\beta}}{\sqrt{m_\beta}}A_\beta(1)-
	\sqrt{m_\beta}A_\beta(0)\right]\right]\!.\nonumber
\end{eqnarray}
It should be noted, however, that not $\lambda$, but the heat conductivity 
coefficient $\chi$ is measured experimentally. It is mutually connected with 
$\lambda$ by the relation $\chi=\lambda/(\rho C_p)$, where $C_p$ is the heat 
capacity at constant pressure.

\noindent
Diffusion velocity in the first approximation reads:
\[
{\mathbf V}_\alpha^{\mathrm d}=-D^\alpha_{\mathrm T}\mbox{\boldmath$\nabla$}
	\ln T-\sum\nolimits_{\beta=1}^MD^{\alpha\beta}{\mathbf d}_\beta,
\]
where 
\begin{equation}
D^\alpha_{\mathrm T}=\sqrt{\frac{k_{\mathrm B}T}{2m_\alpha}}A_\alpha(0)
\label{e8}
\end{equation}
is the thermal diffusion coefficient of a mixture, while
\begin{equation}
D^{\alpha\beta}=-n\sqrt{\frac{k_{\mathrm B}T}{2m_\alpha}}E_{\alpha\beta}(0)
\label{e9}
\end{equation}
is the mutual diffusion coefficient. Quantities $B_{\alpha}(0)$ in 
(\ref{e6}), $A_{\alpha}(0)$ and $A_{\alpha}(1)$ in (\ref{e7}) and 
(\ref{e8}), $E_{\alpha\beta}(0)$ in (\ref{e9}) are nothing but coefficients 
of expansion in the Sonine-Laguerre polynomials. Their general definition 
for an arbitrary potential of interaction is given in \cite{5}. The 
calculations for a special case of a two-component mixture is performed in 
\cite{7}. All the obtained quantities $B_{\alpha}(0)$, $A_{\alpha}(0)$, 
$A_{\alpha}(1)$ and $E_{\alpha\beta}(0)$ ultimately depend on the so-called 
$\mathit{\Omega}$-integrals.

\begin{figure*}[htb]
\begin{centering}
\fbox{\includegraphics*[bb=59 53 547 622,%
	angle=-90,width=15.6cm]{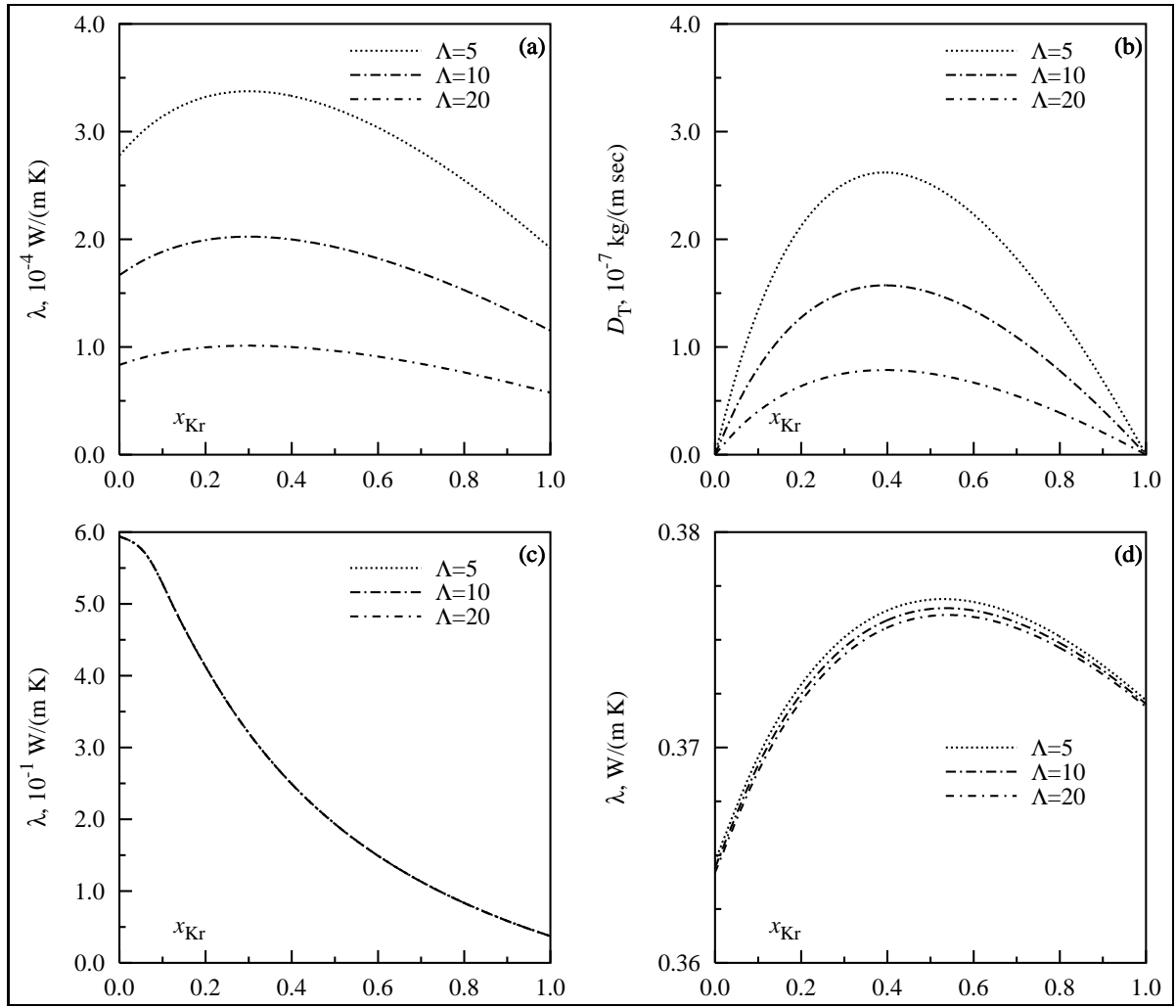}}\\ [1ex]
\end{centering}
\caption{The same as in Fig.~1, but for systems of neutral and 
charged point-like particles at $T=5000$ K for (a), (b) and (c) and at 
$T=6500$ K for (d). Parameter $\Lambda$ indicates for what value of Coulomb 
logarithm calculation was performed.}
\label{f2}
\end{figure*}

Numerical calculation for transport coefficients $\kappa$, $\eta$, 
$\lambda$, $D_{\mathrm T}^\alpha$ and $D^{\alpha\beta}$ has been performed. 
For two- and three-component mixtures of neutral and charged hard spheres 
we studied the dependences of transport coefficients on density, 
temperature, and concentration ratio of some mi\-xtu\-re components 
\cite{18}. There are a lot of approaches which allow to calculate viscosity 
with sufficient accuracy. However, these approaches do not allow to 
calculate well thermal conductivity for dense and moderately dense systems. 
Our theory is devoid of such a circumstance. This is re\-a\-ched due to the 
following: firstly, the Enskog-Landau kinetic equation is obtained by means 
of the non-equilibrium statistical operator method from the first principles 
of statistical mechanics without phe\-no\-me\-no\-lo\-gi\-cal assumptions;
\begin{figure*}[htb]
\begin{centering}
\fbox{\includegraphics*[bb=62 54 547 622,%
	angle=-90,width=15.6cm]{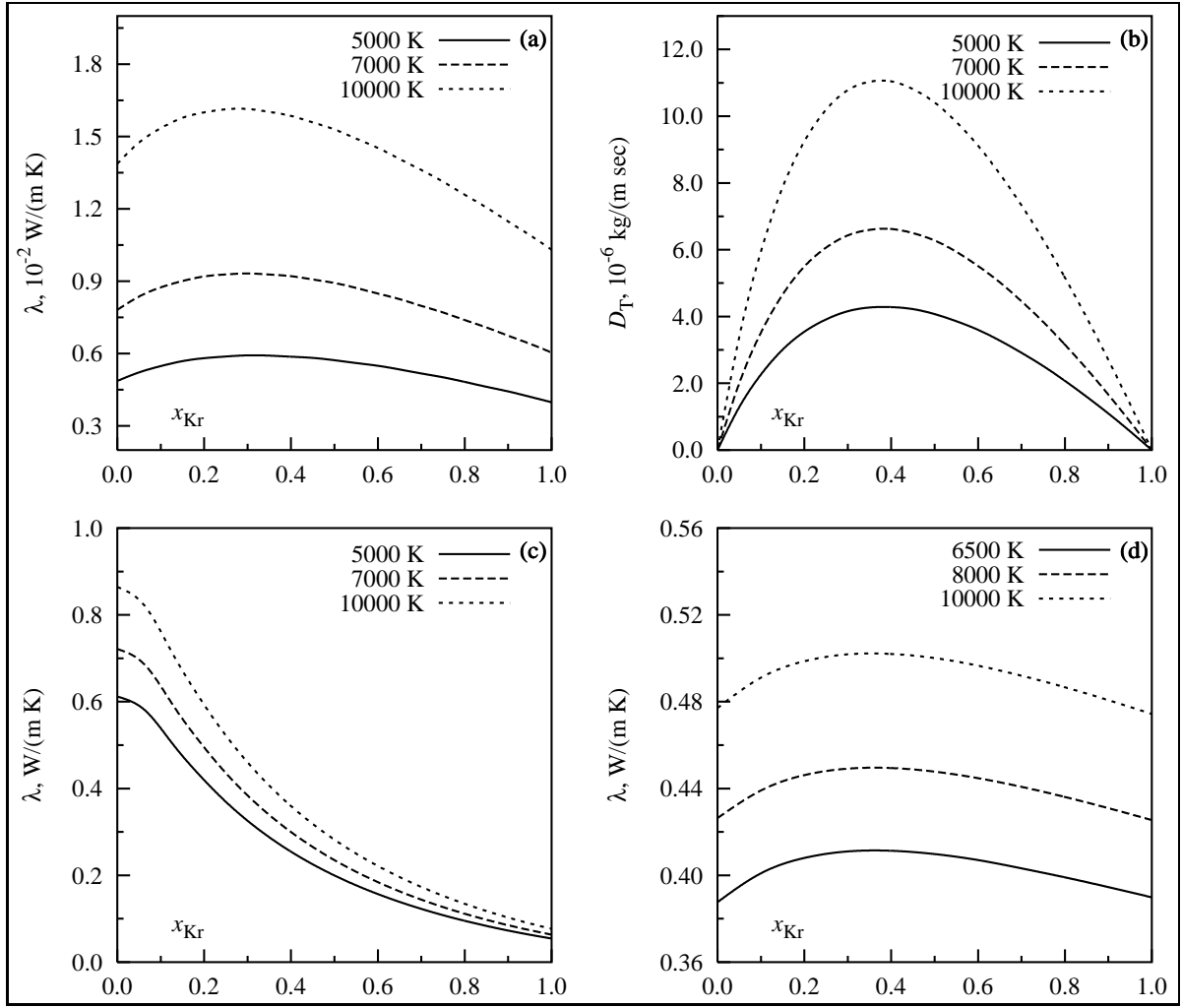}}\\ [1ex]
\end{centering}
\caption{Concentration ratio dependences of transport coefficients of some 
binary and ternary systems of neutral and charged hard spheres. Thermal 
conductivity (a) and thermal diffusion coefficient (b) of a system 
Ar$^+$--Kr$^+$ at total concentration $n=2\cdot10^{20}$ cm$^{-3}$. Thermal 
conductivity of a system He--Kr--Ar$^+$ (c) at total concentration 
$n=2\cdot10^{21}$ cm$^{-3}$, $x_{\mathrm{Ar}}=0.2$ or $n_{\mathrm 
{Ar}}=4\cdot10^{20}$ cm$^{-3}$. Thermal conductivity of a system 
He--Ar$^+$--Kr$^+$ (d) at total concentration $n=1.25\cdot10^{21}$ 
cm$^{-3}$, $x_{\mathrm{He}}=0.6$.}
\label{f1}
\end{figure*}
secondly, kinetics and hy\-dro\-dy\-na\-mics in the studied systems are 
considered simultaneously. The last factor is very important for dense and 
moderately dense systems \cite{13,14,15,16,17}. As a result, we obtain a 
good agreement between the theory and experimental data. In Fig.~\ref{f1} we 
present the concentration ratio dependences of some transport coefficients 
for two- and three-component mixtures for charged hard spheres. In all 
cases, the total concentration is considered to be constant. 
Fig.~\ref{f1}~(b) illustrates the limiting cases for the thermal diffusion 
coefficient $D_{\mathrm T}^\alpha$ of a system Ar$^+$--Kr$^+$ when 
$x_{\mathrm{Kr}}\to0$ or $x_{\mathrm{Kr}}\to1$, that is for one-component 
systems. In this case, diffusion thermodynamic forces vanish and thermal 
diffusion vanishes too. Fig.~\ref{f2} shows numerical calculations for the 
same systems as in Fig.~\ref{f1}, but for point-like neutral and charged 
particles. Transport coefficients for last ones were calculated as for the 
usual Boltzmann-Landau kinetic equation \cite{11}. The magnitude of 
transport coefficients therewith slows down. It is interesting to note that 
in three-component systems of point-like particles (Fig.~\ref{f2} (c) and 
(d)) transport coefficients depend on $\Lambda$ slightly. It is well to bear 
in mind that this takes place if at least one component is not charged.

From the present letter, one can draw the following conclusions. The obtained
Enskog-Landau kinetic equation for charged hard spheres turned out to be 
very useful for several purposes. First of all, the collision integral of 
this equation does not contain a divergency at small distances. Secondly, 
the normal solution and {\slshape all} transport coefficients have 
analytical structure. They can be easily used to study some specific 
systems. Finally, the analytical structure of transport coefficients allows 
us to find fast and easily systems, which can be best described by the 
obtained kinetic equation, as well as density and temperature ranges, where 
the agreement between the theory and experimental data is the closest.
At the same time, our theory has not met with success in treatment of the free 
electrons role. Basing on the conclusions by Ichimaru {\slshape et. al.} 
\cite{19}, consistent treatment of electrons is possible only within the 
frame of quantum kinetic theory. Our theory is purely classical. Just the 
same, our papers \cite{1,7} show very good agreement between theoretical 
calculations and experimental data.

The next step in this theory is to calculate a dynamical screening radius in 
a system. Partially this problem has been already solved in our recent 
paper \cite{20}.


\begin{thebibliography}{99}
\bibitem{1} A.E. Kobryn, V.G. Morozov, I.P. Omelyan, M.V. To\-kar\-chuk, 
	Phy\-si\-ca A, {\bf 230}, 189 (1996).
\bibitem{2} D.N. Zubarev, V.G. Morozov, I.P. Omelyan, M.V. To\-kar\-chuk, 
	Teor. Mat. Fiz., {\bf 87}, 113 (1991).
\bibitem{3} A.E. Kobryn, I.P. Omelyan, M.V. Tokarchuk, 
	Phys. Lett. A, {\bf 223}, 37 (1996).
\bibitem{4} A.E. Kobryn, I.P. Omelyan, M.V. Tokarchuk, 
	Cond. Matt. Phys. Issue {\bf 8}, 75 (1996).
\bibitem{5} J.H. Ferziger, H.G. Kaper, {\itshape Mathematical theory of 
	transport processes in gases} (North-Holland, Amsterdam, 1972).
\bibitem{6} D.N. Zubarev, A.D. Khonkin, 
	Teor. Mat. Fiz., {\bf 11}, 403 (1972).
\bibitem{7} A.E. Kobryn, I.P. Omelyan, M.V. Tokarchuk, Physica A, 
	{\bf 268}, 607 (1999).
\bibitem{8} A.V. Maximov, V.P. Silin, M.V. Chegotov, Fizika Plazmy,
	{\bf 16}, 575 (1990).
\bibitem{9} V.P. Silin, {\itshape Introduction to the kinetic theory of 
	gases} (Nauka, Moscow, 1971).
\bibitem{10} J.~van~de~Ree, Physica, {\bf 36}, 118 (1967).
\bibitem{11} Yu.L. Klimontovich, {\itshape Kinetic theory of nonideal gas 
	and nonideal plasmas} (Pergamon, Oxford, 1982).
\bibitem{12} R. Balescu, {\itshape Transport processes in plasmas}. Vol. 1, 
	{\itshape Classical transport} (North Holland, Amsterdam, 1988).
\bibitem{13} Yu.L. Klimontovich, Phys. Lett. A, {\bf 170}, 434 (1992).
\bibitem{14} Yu.L. Klimontovich, Teor, Mat. Fiz., {\bf 92}, 312 (1992).
\bibitem{15} D.N. Zubarev, V.G. Morozov, I.P. Omelyan, M.V.To\-kar\-chuk, 
	Te\-or. Mat. Fiz., {\bf 96}, 325 (1993).
\bibitem{16} M.V. Tokarchuk, I.P. Omelyan, A.E. Kobryn, 
	Cond. Matt. Phys. {\bf 1}, 687 (1998).
\bibitem{17} D.N. Zubarev, V.G. Morozov, G. R\"opke, {\itshape Statistical 
	mechanics of nonequilibrium processes}. Vol. 1, {\itshape Basic 
	concepts, kinetic theory} (Academie Verlag, Berlin, 1996).
\bibitem{18} M.V. Tokarchuk, A.E. Kobryn, Y.A. Humenyuk, Preprint of the 
	Institute for Condensed Matter Physics, ICMP-99-08U at
	www.icmp.lviv.ua/icmp/preprints/ps/9908Ups.gz.
\bibitem{19} S. Ichimaru, H. Iyetomi, S. Tanaka, Phys. Rep., {\bf 149}, 91 
	(1987).
\bibitem{20} A.E. Kobryn, I.P. Omelyan, M.V. Tokarchuk, J. Stat. Phys., 
	{\bf 92}, 973 (1998).
\end{thebibliography}
\end{document}